\documentclass[nohyper,notoc]{JHEP}
\input epsf
\usepackage{epsfig}

\newcommand{\mapright}[1]{\smash{\mathop{
\hbox to 0.2cm{\rightarrowfill}}\limits_{#1}}}

\title{ Lattice monopole action in pure SU(3) QCD }

\author{Kentarou Yamagishi and Tsuneo Suzuki \\ 
Institute for Theoretical Physics, Kanazawa University, \\
Kanazawa 920-1192, Japan.\\ 
E-mail: \email{yamaken@hep.s.kanazawa-u.ac.jp}\\
E-mail: \email{suzuki@hep.s.kanazawa-u.ac.jp}}

\author{Shun-ichi Kitahara \\
Jumonji University, Niiza, Saitama
352-8510, Japan.\\
E-mail: \email{kitahara@jumonji-u.ac.jp}}

\abstract{
We obtain an almost perfect monopole action 
numerically after abelian projection
in pure SU(3) lattice QCD.\ 
Performing block-spin transformations on the dual lattice,\ 
the action fixed depends only on a physical scale $b$.\
Monopole condensation occurs for large $b$ region.\ 
The numerical results show that two-point monopole interactions
are dominant for large $b$.\ We next perform the block-spin
transformation analytically in a simplified case of two-point monopole
interactions with a Wilson loop on the fine lattice.\
The perfect operator evaluating the static
quark potential on the coarse $b$-lattice
are derived.\  The monopole partition function can be transformed into
that of the string model.\ The static potential and the string tension 
are estimated in the string model framework.\ 
The rotational invariance of the static potential is recovered,\ 
but the string tension is a little larger than the physical one.
}


\preprint{hep-lat/0002001 \\
KANAZAWA-99-27}

\begin{document}

\section{Introduction}
The quark confinement is a key problem to understand non-perturbative
phenomena of QCD.\ The dual Meissner effect is a promising
candidate [1].\ 't\ Hooft suggested the idea of abelian projection
[2].\ $SU(N)$ QCD is reduced to a $U(1)^{N-1}$ abelian gauge theory
with magnetic monopoles by a partial gauge fixing (abelian
projection).\ 
The dual Meissner effect is caused by the monopole
condensation.\ 
Monopoles are responsible for the confinement as in compact QED [3-8].\par
Monte-Carlo simulations of lattice QCD are a most powerful
method to study the non-perturbative phenomena.\ 
Numerical studies of the abelian projection in the maximally abelian
(MA) gauge have confirmed the 't\ Hooft conjecture.\
In MA gauge [9,10],\ the string tension derived from abelian Wilson loops
gives almost the same value as that of non-abelian Wilson loop
(abelian dominance) [11-13].\
Moreover,\ the monopole
contribution to the abelian Wilson loops alone reproduces the string 
tension in SU(2) QCD (monopole dominance) [14-16].\ 
The abelian and the monopole dominances are seen also in the behavior of 
the Polyakov loop in $T\ne 0$ SU(2) and SU(3) QCD [17-19].\
These results support 
the 't\ Hooft conjecture.\par
Wilson's idea of a block-spin transformation and 
a renormalized trajectory (RT) are useful when we study the
continuum limit on available lattices [20].\
The lattice action on RT has no lattice artifact 
and hence it is called quantum perfect action.\
It reproduces the same physical results 
as in the continuum limit.\ 
It is challenging to get the perfect lattice action for the infrared
region of QCD.\
For that purpose,\ we have to extract a dynamical variable which 
plays a dominant role in 
the infrared region.\ 
The numerical evidence for the monopole dominance
suggests that low-energy QCD can be
described by an effective action on the dual lattice in terms of a
dual quantity like monopole currents.\ 
It is very interesting to obtain such an effective monopole action.\
Also it is challenging to set the perfect monopole action with the
help of block-spin transformations for monopole currents.\
The monopole action can be obtained numerically
by an inverse Monte-Carlo method [21].\
A block-spin transformation on the dual lattice
can be performed by considering an $n$-blocked monopole current [22].\ 
The detailed study of the effective monopole action in pure SU(2) QCD has
been done [23-25].\ The monopole action determined in SU(2) has the 
following important features:
\begin{enumerate}
\item
Two-point interactions are dominant in the infrared region and
coupling constants decrease rapidly as the distance between two
monopole currents increases.
\item
The action fixed seems to satisfy the scaling behavior,\ that is,\ 
it depends only on a physical scale $b=na(\beta)$.\ 
This suggests that the action is near to RT.
\item
Monopole condensation seems to occur for large $b$ region from the
energy-entropy balance.
\end{enumerate}
\par
In order to test the validity of the statement that
the action is near to RT,\ we need to check the
restoration of the continuum rotational invariance.\
Then,\ we have to determine first the correct forms of physical
operators (quantum perfect operators) on the blocked lattice.\
In the above pure SU(2) study,\
Fujimoto {\it et al.}\ [26] have taken the following steps:
\begin{enumerate}
\item
Study the renormalization flow on the projected space of
two-point monopole interactions alone,\ considering that
they are dominant numerically in the infrared region.\
\item
The static potential between quark and antiquark can be
evaluated by the expectation value of the Wilson loop in the
continuum limit.\
Hence the Wilson loop can be regarded as the correct operator
evaluating the static potential on the fine $a$-lattice also.\
\item
Perform the block spin transformation {\it analytically},\
starting from the two-point monopole interactions with the
Wilson loop on the fine $a$-lattice.\
The quantum perfect action and the quantum perfect operator
on the coarse $b$-lattice can be obtained {\it analytically}
when we take the $a\to 0$ limit for fixed $b=na$.\
\item
Compare the quantum perfect action composed of
two-point monopole interactions with the effective action
numerically determined from the inverse Monte-Carlo method.\
The parameters in the perfect monopole action 
and the quantum perfect operator are fixed then.\
\item
The effective action with the quantum perfect operator
can be transformed into that of the string model.\
Since the strong-coupling expansion is found to work well
in the string model,\ we see that the static potential is estimated
analytically by the classical part alone and the continuum rotational
invariance is restored.\
\end{enumerate}
It is not so straightforward to extend these studies to pure SU(3) QCD,\ since
there are three monopole currents $k_\mu^{(a)} (s)$ satisfying one
constraint $\sum_{a=1}^3 k_\mu^{(a)} (s)=0$.\ 
So far,\ SU(3) monopole action composed of only one kind of monopole current
after integrating out the other two has been derived 
in the case of two-point interactions [27].\ 
In this case,\ the same method can be applied as done in pure SU(2) QCD and 
monopole condensation seems to occur from the energy-entropy balance
in rather strong coupling region.\ But the scaling seen in the pure SU(2) case
was not seen.\
Also it is important to study effective action for two independent
monopole currents in order to see the characteristic features of 
pure SU(3) QCD.\par 
The purpose of this paper is to report the results of the extensive studies
of SU(3) monopole actions,\ especially in terms of two independent
monopole currents.\
In Section 2 we briefly review the SU(3) monopole current on the lattice
and the inverse Monte-Carlo method.\ 
The numerical results of the SU(3) monopole action are shown in Section
3.\ In Section 4 we perform a block-spin transformation of the monopole
current analytically,\ considering only the infrared dominant
two-point monopole interactions.\
In Section 5,\
transforming the monopole action into that of the string model,\ 
we calculate the static potential analytically.\
In Section 6 the
string tension is estimated from the results of previous Sections.\
The conclusions are given in Section 7.\
The correspondence between the monopole action and
the dual abelian Higgs theory 
(dual Ginzburg-Landau theory) 
is given in Appendix.\

\section{Monopole current 
and the inverse Monte-Carlo method}
We extract a $U(1)^2$ link field 
$u_\mu(s)$ from a SU(3) link
field $U_\mu(s)$ after abelian projection called MA gauge.\
In the SU(2) case,\ a $U(1)$ link field is defined as 
$u_\mu (s) =\rm{diag}(e^{i\theta_{\mu}^{(1)}},e^{i\theta_{\mu}^{(2)}})$,\
where $\theta_\mu^{(a)}(s)\equiv 
\arg [U_\mu(s)]_{aa}\ (a=1,2)$.\
It satisfies $\det(u_\mu(s))=1$ due to $\sum_{a=1}^2
\theta_{\mu}^{(a)} = 0$.\
Since $\sum_{a=1}^3 \theta_{\mu}^{(a)} \ne 0$ in the SU(3)
case,\ the definition of $U(1)^2$ link field necessarily becomes
more complicated.\ 
In this study,\ we use the definition of Ref.\ [28,29].\footnote{
Another definition is seen in Ref.\ [10].\
Both definitions are equivalent in the continuum limit.}
The fields transforming as photon fields under $U(1)^2$
are defined as follows:
\begin{equation}
\theta_{\mu}^{(a)} \equiv {\rm{arg}}[U_{\mu}]_{aa} -
\frac13\phi_{\mu},\\\\\\\
\left.\phi_{\mu} \equiv \sum_{a=1}^3 {\rm{arg}}[U_{\mu}]_{aa}
\right|_{{\rm{mod}}
2\pi}\ \in [-\pi,\pi).
\end{equation}
The $U(1)^2$ link field defined by
$u_\mu (s) = \rm{diag}(e^{i\theta_{\mu}^{(1)}},e^{i\theta_{\mu}^{(2)}}
,e^{i\theta_{\mu}^{(3)}})$ satisfies  $\det(u_\mu(s))=1$.\par
If the $U(1)^2$ field strength is defined as
$\overline{\Theta}_{\mu\nu}^{(a)} \equiv
\partial_{\mu} \theta_{\nu}^{(a)}
- \partial_{\nu} \theta_{\mu}^{(a)} \  (\rm{mod} 2\pi)$,\ then
$\sum_{a}\overline{\Theta}_{\mu\nu}^{(a)}
=2\pi l\ (l=0,\pm1)$ and is not always zero.\ 
When $l=+1(-1)$,\ $\overline{\Theta}_{\mu\nu}^{(a)}$ is redefined.\ 
If $\overline{\Theta}_{\mu\nu}^{(a)}$ is the
maximum(minimum) of $(\overline{\Theta}_{\mu\nu}^{(1)}
,\overline{\Theta}_{\mu\nu}^{(2)},
\overline{\Theta}_{\mu\nu}^{(3)})$,\ it is redefined as
$\overline{\Theta}_{\mu\nu}^{(a)} - 2\pi(+2\pi)$.\ Others do not change.\
Then new $U(1)^2$ field strengths satisfy
$\sum_a\overline{\Theta}_{\mu\nu}^{(a)} =0$.\ 
Monopole currents are defined as
$k_{\mu}^{(a)} \equiv  1/4\pi\epsilon_{\mu\nu\rho\sigma} \partial_{\nu}
\overline{\Theta}_{\rho\sigma}^{(a)}$ by DeGrand-Toussaint(D-T) [7]
which satisfies the constraint
\begin{equation}
\sum_a k_{\mu}^{(a)} = 0.
\end{equation}
\ We want to get an effective monopole action on the dual lattice integrating
out the degrees of freedom except for the monopole currents:
\begin{eqnarray}
Z&=&\int {\cal D}U\delta(X^{off})\Delta_F (U)e^{-S(U)}\\
&=&\int {\cal D}u [\int {\cal D}c \delta(X^{off})\Delta_F (U)e^{-S(U)}]
\\
&=&\int {\cal D}u e^{-S_{eff}(u)}\\
&=&\sum_{k^{(a)} \in Z}\delta_{\partial'_\mu k_\mu^{(a)},0}
\delta_{\Sigma_a k^{(a)},0}
\int {\cal D}u \delta(k^{(a)},u)e^{-S_{eff}(u)}\\
&=&\sum_{k^{(a)} \in Z}\delta_{\partial'_\mu k_\mu^{(a)},0}
\delta_{\Sigma_a k^{(a)},0}
e^{-S[k^{(a)}]},
\end{eqnarray}
where $U_{\mu}(s)=c_\mu(s)u_\mu(s)$ and
$X^{off}$ are the off-diagonal part of the following quantity:
\begin{equation}
X(s)=\sum_{\mu,a}[U_\mu (s)\Lambda_a U_\mu^{\dagger}(s)
+U_\mu^{\dagger} (s-\hat\mu)\Lambda_a U_\mu(s-\hat\mu),\ \Lambda_a],
\end{equation}
\begin{eqnarray}
\Lambda_1&=&
\left(
\begin{array}{ccc}
1 & 0 & 0 \\
0 & -1& 0 \\
0 & 0 & 0
\end{array}
\right),\
\Lambda_2=
\left(
\begin{array}{ccc}
-1 & 0 & 0 \\
0 & 0 & 0 \\
0 & 0 & 1
\end{array}
\right),\ 
\Lambda_3=
\left(
\begin{array}{ccc}
0 & 0 & 0\\
0 & 1 & 0\\
0 & 0 & -1
\end{array}
\right).
\end{eqnarray}
$\Delta_F(U)$ is the Faddeev-Popov determinant and $\delta 
(k^{(a)},u)$ is the delta function corresponding to
the D-T definition of the monopole current.\par
We can perform the above integration numerically.\
We create vacuum ensembles of monopole currents\ 
$\{k_{\mu}^{(a)}(s)\}$
using the Monte-Carlo method and the above definition of the monopole current.\
Then we derive the effective monopole action 
using the Swendsen method [21] which is one of the inverse Monte-Carlo methods.\par
In order to take the continuum limit,\ we perform the block-spin transformation
on the dual lattice
by defining an $n$-blocked monopole current [22]:
\begin{eqnarray}
K_{\mu}^{(a)}(s^{(n)})&\equiv&\sum_{i,j,m=0}^{n-1} k_{\mu}^{(a)}
(ns+(n-1)\hat{\mu}+i\hat{\nu}+j\hat{\rho}+m\hat{\sigma}).
\end{eqnarray}
If the action numerically obtained satisfies a scaling behavior
$S[K^{(a)},n,a(\beta)]=S[K^{(a)},b=na(\beta)]$,\ that is,\ 
the action depends only on 
$b$,\ 
the continuum limit can be taken as $n\to\infty,\ a(\beta)\to 0$
for a fixed physical length $b=na(\beta)$.\par
The original Swendsen method must be extended due to 
the conservation law of the monopole current 
$\partial'_{\mu}k_{\mu}^{(a)}(s)=0$ as in the SU(2) case [23].\
Let us assume the form of the monopole action as 
$S[k^{(a)}]= \sum_iG_i S_i[k^{(a)}]$ ($G_i$ is the coupling 
constant for the operator $S_i[k^{(a)}]$) and define
${\tilde{S}}_i [k^{(a)}]$ as a part of 
$S_i [k^{(a)}]$ which contains the currents around a
specific plaquette $(s',\hat\mu',\hat\nu')$.\ When we consider the
expectation value of some operator\ $O_i[k^{(a)}]$,\
the following identity holds as in
the SU(2) case:
\begin{equation}
\langle O_i[k^{(a)}]\rangle=\langle\overline{O}_i[k^{(a)}]\rangle,
\end{equation}
\begin{equation}
\langle O_i[k^{(a)}]\rangle \equiv
\frac{\sum_{k^{(a)} \in Z}\delta_{\partial'_\mu k_\mu^{(a)},0}
\delta_{\Sigma_a k^{(a)},0}
O_i[k^{(a)}]\exp(-\sum_j G_j S_j[k^{(a)}])}
{\sum_{k^{(a)} \in Z}\delta_{\partial'_\mu k_\mu^{(a)},0}
\delta_{\Sigma_a k^{(a)},0}
\exp(-\sum_j G_j S_j[k^{(a)}])},
\end{equation}
\begin{equation}
\overline{O}_i[k^{(a)}] \equiv
\frac{\sum_{M^{(a)} \in Z}
\delta_{\Sigma_a M^{(a)},0}
O_i[k'^{(a)}]\exp(-\sum_j G_j \tilde{S}_j[k'^{(a)}])}
{\sum_{M^{(a)} \in Z}
\delta_{\Sigma_a M^{(a)},0}
\exp(-\sum_j G_j \tilde{S}_j[k'^{(a)}])},
\end{equation}
\begin{equation}
{k'}_{\mu}^{(a)}(s) \equiv
k_{\mu}^{(a)}(s)+M^{(a)}(\delta_{s,s'}\delta_{\mu,
\mu'}
+\delta_{s,s'+\hat{\mu}'}\delta_{\mu,\nu'}-\delta_{s,s'+\hat{\nu}'}
\delta_{\mu,\mu'}-
\delta_{s,s'}\delta_{\mu,\nu'}).
\end{equation}
We use this identity to determine the values of the couplings
$G_i$ iteratively.\   
Choose trial couplings {$\tilde{G}_i$} suitably.\ If
$\tilde{G}_i$ are not equal to $G_i$
for all $i$,\ then
$\Delta\tilde{G}_i \equiv G_i-\tilde{G}_i$ are estimated
from the expansion of Eq.\ (2.11) up to the first order of 
$\Delta\tilde{G}_i$:
\begin{equation}
\langle O_i-\overline{O}_i\rangle \cong
\langle\overline{O}_i\overline{S}_j -
\overline{O_i S_j}\rangle
\Delta\tilde{G}_j.
\end{equation}
$\tilde{G'}_i=\tilde{G}_i+\Delta\tilde{G}_i$ are used
as the next trial couplings.\
These procedures continue iteratively until the couplings
converge.

\section{Numerical results}

Practically,\ we must truncate the form of the monopole action to derive
it numerically.\
We know that short-distant and two-point
interactions are dominant in the SU(2) case.\
Here,\ we assume the following form of the SU(3) monopole action:
\begin{eqnarray}
S[k^{(a)}]&=&\sum_iG_i
(S_i[k^{(1)}]+S_i[k^{(2)}]+S_i[k^{(3)}])\\
&=&\sum_i G_i(S_i[k^{(1)}]+S_i[k^{(2)}]+S_i[-k^{(1)}-k^{(2)}]).
\end{eqnarray}
The Weyl symmetry (the species permutation
symmetry of the monopole currents) remains after abelian projection.\
We adopt 27 two-point interactions whose distances are up to
$3na(\beta)$ [23] and 4,\ 6-point interactions of the  following simple form:
\begin{equation}
\sum_{a,s}\{\sum_{\mu=-4}^4({k}_{\mu}^{(a)} (s))^2\}^2,
\quad\quad
\sum_{a,s}\{\sum_{\mu=-4}^4({k}_{\mu}^{(a)} (s))^2\}^3.
\end{equation}
The lattice size is $48^4$ from
$\beta=5.6$ to $\beta=6.4$.\ After thermalization,\ 30
configurations are used for the average.\ 
We determine the lattice spacing $a(\beta)$ without using the
theoretical asymptotic
beta function.\ It is given by the relation
$a(\beta)=\sqrt{\sigma(\beta)/\sigma_{ph}}$ [30],\ where $\sigma_{ph}$ is the
physical string tension 
\footnote{If we use $\sigma_{ph}^{1/2}=440$\ [MeV],\ 
$b=2\sigma_{ph}^{-1/2}$ is approximately equal to 0.9 [fm].}.\
The results are summarized as follows:
\begin{enumerate}
\item
The monopole action for two independent types of monopole currents is obtained
clearly (Figure 1).\ 
The qualitative behaviors are the same as in the SU(2) monopole
action:\ 

\begin{figure}[htb]
\begin{minipage}{72mm}
\epsfxsize=7.2cm
\epsfysize=7.2cm
\begin{center}
\leavevmode
\epsfbox{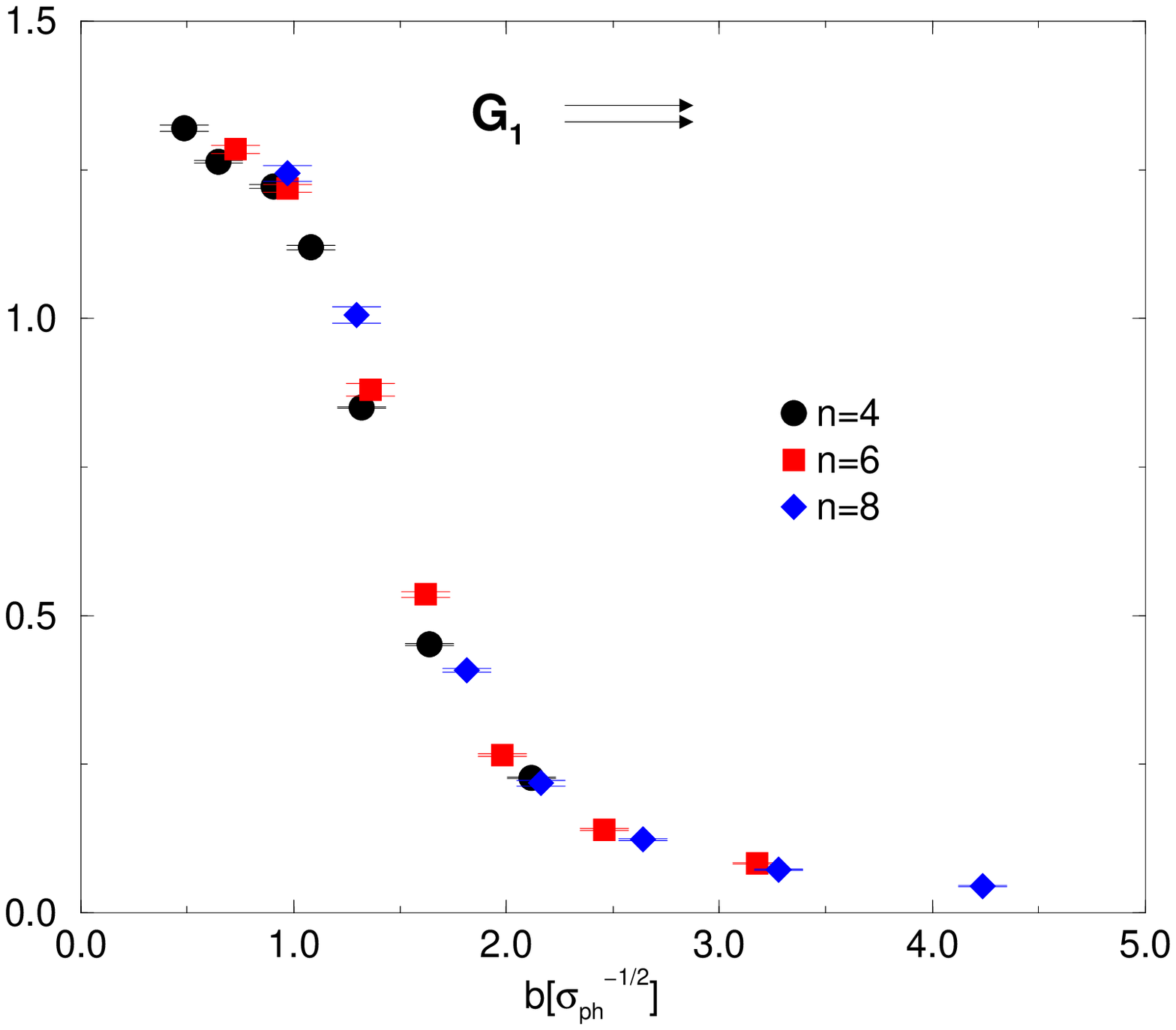}
\end{center}
\label{fig:9}
\end{minipage}
\hfill
\begin{minipage}{72mm}
\epsfxsize=7.2cm
\epsfysize=7.2cm
\begin{center}
\leavevmode
\epsfbox{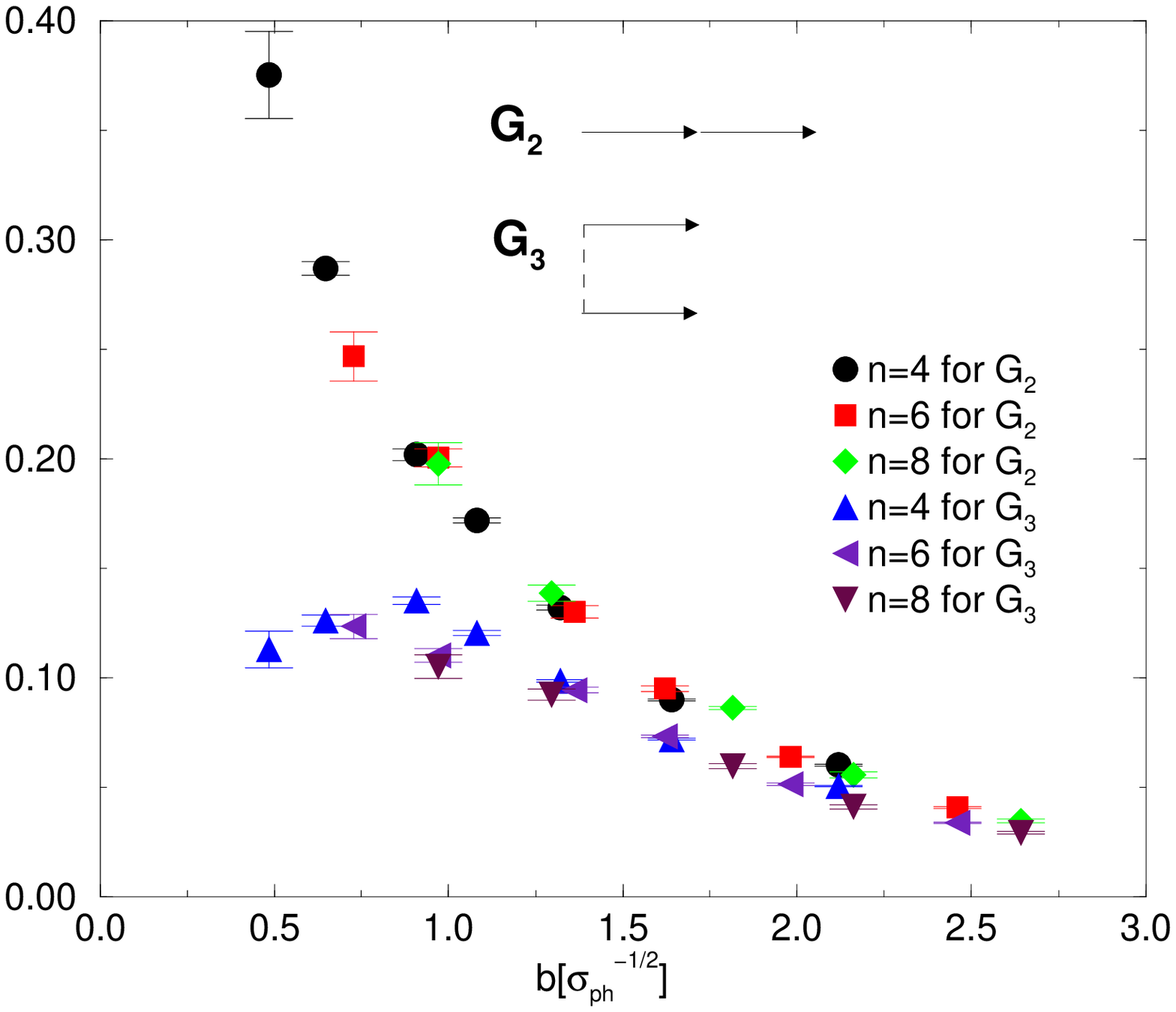}
\end{center}
\label{fig:9}
\end{minipage}

\begin{minipage}{72mm}
\epsfxsize=7.2cm
\epsfysize=7.2cm
\begin{center}
\leavevmode
\epsfbox{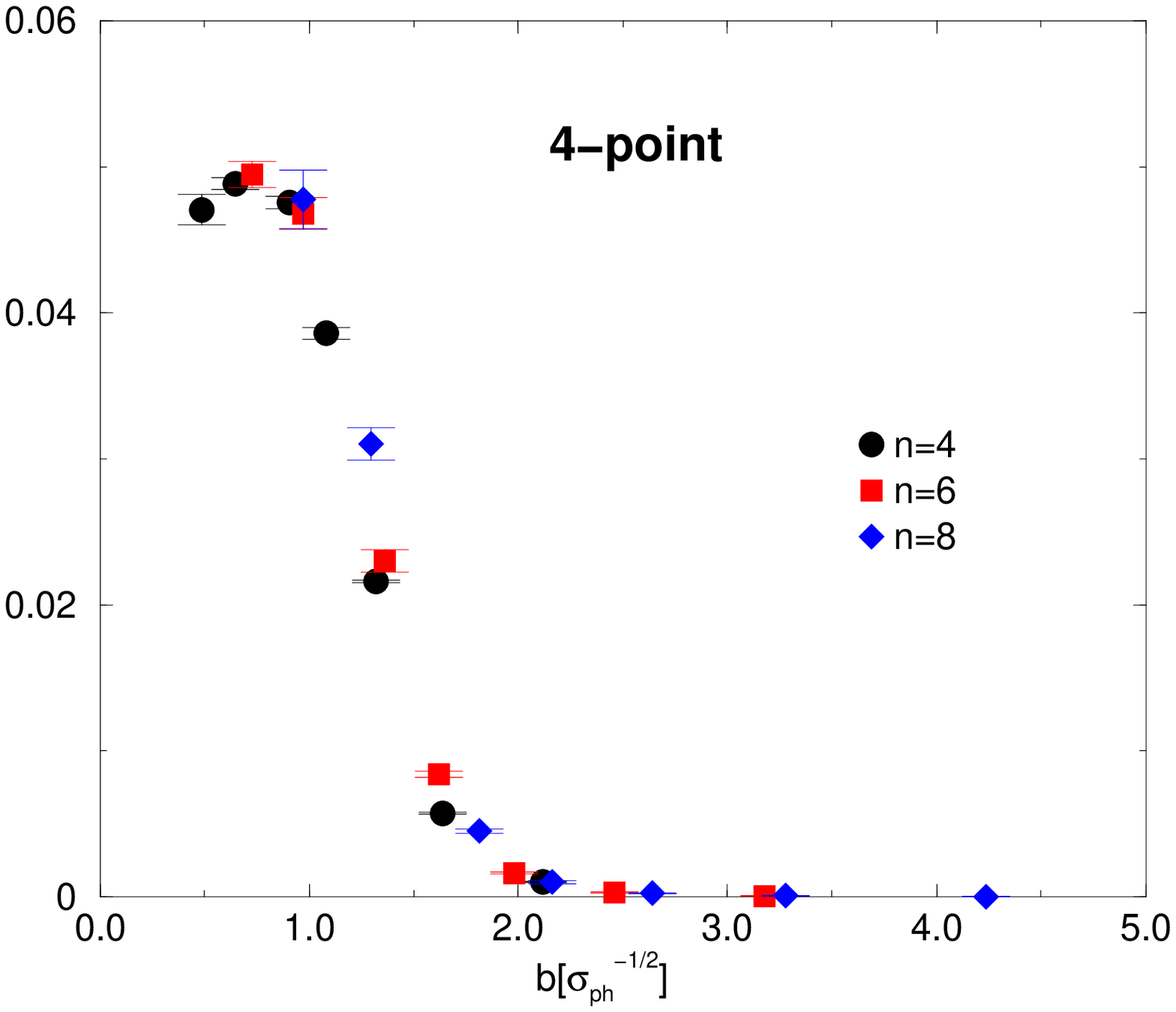}
\end{center}
\label{fig:9}
\end{minipage}
\hfill
\begin{minipage}{72mm}
\epsfxsize=7.2cm
\epsfysize=7.2cm
\begin{center}
\leavevmode
\epsfbox{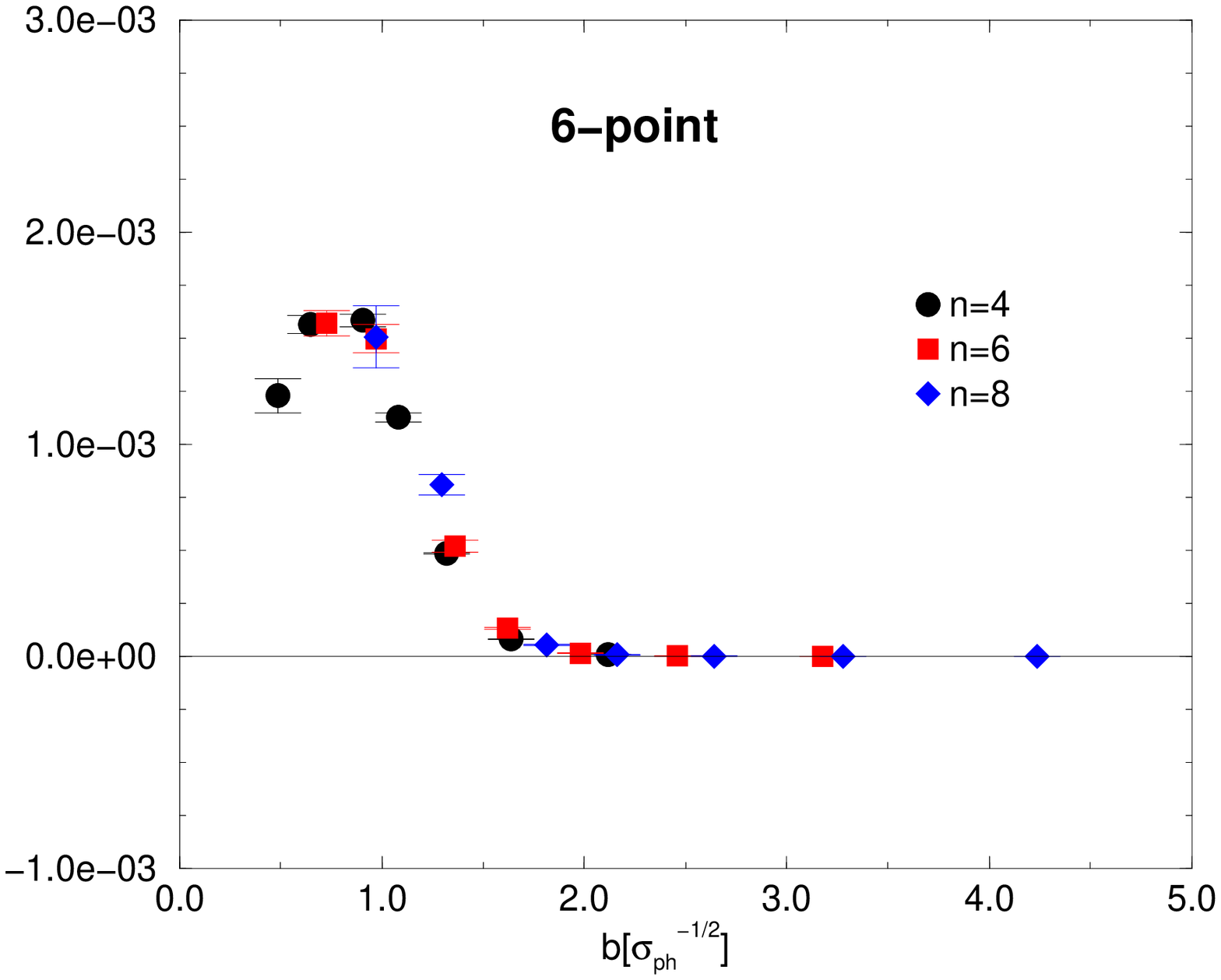}
\end{center}
\label{fig:9}
\end{minipage}
\caption{
The coupling $G_i$ versus $b$ in two-current case.\
The 4pt couplings are negative really.
}
\end{figure}

\begin{enumerate}
\item
The monopole action has a compact form.\
The self-interaction \par
$G_1 \sum_{a,s,\mu}(k_{\mu}^{(a)}(s))^2$
is dominant and the coupling constants $G_i$ decrease
rapidly as the distance between the two monopole currents increases.\
\item
The coupling constants have a direction dependence which
is expected after blocking.\
Two nearest-neighbor interactions \par 
$G_2 \sum_{a,s,\mu}k_{\mu}^{(a)} (s)k_{\mu}^{(a)} (s+\hat\mu)$ and 
 $G_3 \sum_{a,s,\mu\ne\nu}
k_{\mu}^{(a)} (s)k_{\mu}^{(a)} (s+\hat\nu)$\par
are quite different
for small $b$ region.\
\item
The coupling constants $G_i$ become very small for large $b$ region.\
\item
The simple 4,\ 6-point interactions become negligibly small
for large $b > 2\sigma_{ph}^{-1/2}$ .\
Two-point interactions are relatively dominant
for large $b$ region.\ 
\item
The scaling behavior
holds well for $n=4,\ 6,\ 8$ data,\
if the physical scale $b=na(\beta)$ is taken in unit of the string
tension $\sqrt{\sigma_{ph}}$.\
The action seems to be very near to RT
on which
one can take the continuum limit.
\end{enumerate}

\item

In order to study if monopole condensation occurs by 
energy-entropy balance,\ we derive the monopole action,\
considering only one type of the monopole current.\
For simplicity,\ only two-point interactions are taken into account.\
The scaling was not seen in the previous study where
the two loop perturbation value was used for $a(\beta)$ [27].\ 
When $a(\beta)$ is fixed by the string tension,\ 
the scaling is seen beautifully in Figure 2.\
Since we restrict ourselves to one type of the monopole current,\
the entropy of the monopole current loop is given approximately by
$\ln7 \times L$ ($L$ is a length of one long loop).\
From the previous study,\ we know that only one long loop
and some short loops of monopoles exist in the vacuum and the value
of the action is well approximated as $G_1\times L$.
Figure 2 shows that the entropy dominates
over the energy in the large $b$ region,\ {\it i.e.,}
$G_1(b) < \ln 7$.\
\end{enumerate}

\begin{figure}[tb]
\epsfxsize=10cm
\begin{center}
\leavevmode
\epsfbox{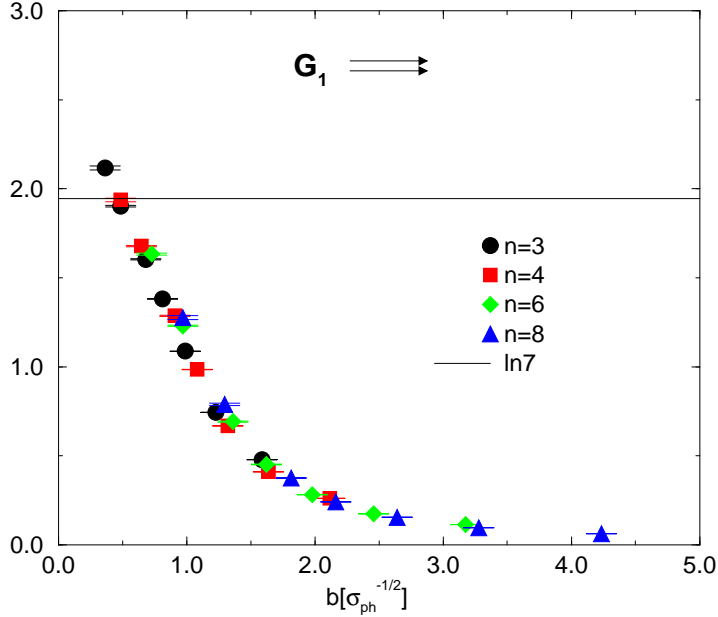}
\end{center}
\caption{The self coupling versus b in the one-current case.
}
\label{fig:5}
\end{figure}

\section{Perfect action and perfect operator for monopole current}

The monopole action seems to satisfy the scaling behavior.\
The scaling is the characteristic of the perfect action.\
The perfect action reproduces the continuum rotational invariance.\
To test the rotational invariance,\ 
we have to determine the form of physical operators on the blocked
lattice.\ For that purpose,\ as done in the SU(2) case,\ 
we perform a block-spin transformation  
from the small $a$-lattice ($a\to 0$)
to the finite $b=na$ lattice
analytically,\ restricting ourselves to a simple case of 
two-point monopole interactions with a monopole Wilson loop.\ 
Note that two-point interactions are dominant for large $b$ region
also in pure SU(3).\par
We start from the following action on the $a$-lattice:
\begin{eqnarray}
Z[J]&=&\sum_{k^{(a)} \in Z}\delta_{\partial'_\mu k_\mu^{(a)},0}
\delta_{\Sigma_a k^{(a)},0}\exp\{-\sum_{a,s,s'}
k_\mu^{(a)}(s)D_0(s-s')k_\mu^{(a)}(s')\nonumber\\
&&+2\pi i\sum_{s}N_\mu(s)k_\mu^{(1)}(s)\},
\end{eqnarray}
\begin{eqnarray}
\ \ \ \ \ \ \ \ \ 
N_{\mu}(s)&=&\sum_{s'}\Delta^{-1}(s-s')\frac12 \epsilon_{\mu\alpha
\beta\gamma}\partial_{\alpha}S_{\beta\gamma}(s'+\hat\mu),\ \ \
{\partial'}_\alpha S_{\alpha\beta}(s)=J_\beta (s).
\end{eqnarray}
Here $J_\beta (s)$ is an electric current around the Wilson loop.\ 
Note that in Eq.\ (4.1),\ the static potential between quark and
antiquark with $a=1$ is considered.\
We have used the monopole Wilson loop operator as in Ref.\ [26].\ 
As the surface $S_{\alpha\beta}$,\ 
we can take any open surface with
the fixed boundary $C$,\
since the monopole Wilson loop operator with the closed surface
is unity due to 4-dimensional linking number in the continuum limit.\
Here we take the flat surface as $S_{\alpha\beta}$.\ 
We expand $D_0 (s)$ and adopt the following first three terms for simplicity:
\begin{eqnarray}
D_0(s-s')&=&\alpha\Delta^{-1}(s-s')+\beta\delta_{s,s'}+\gamma\Delta(s-s').
\end{eqnarray}
The above form of the monopole action is derived from the 
dual Ginzburg-Landau (DGL) theory also (for details,\ see Appendix).\par
We perform the block-spin transformation,\ using the definition of
$n$-blocked monopole currents $K^{(a)}$ in Eq.\ (2.10):
\begin{eqnarray}
Z[K^{(a)},J]&=&\sum_{k^{(a)} \in Z}\delta_{\partial'_\mu k_\mu^{(a)},0}
\delta_{\Sigma_a
k^{(a)},0}
\exp\{-\sum_{a,s,s'}k_\mu^{(a)} (s)D_0(s-s')k_\mu^{(a)} (s')
\nonumber\\
&&+2\pi i \sum_{s}N_\mu (s)k_\mu^{(1)} (s)\}\nonumber\\
&&\times \prod_a \delta(K_\mu^{(a)}(s^{(n)})
-\sum_{i,j,k=0}^{n-1}k_\mu^{(a)}(ns^{(n)}+(n-1)\hat\mu
+i\hat\nu+j\hat\rho+k\hat\sigma))\\
&=&\int_{-\pi}^{\pi}{\cal D}\mbox{\boldmath $\gamma$}
\int_{-\pi}^{\pi}{\cal D}\mbox{\boldmath $B$}\int_{-\pi}^{\pi}
{\cal D}\phi \sum_{k^{(a)} \in Z}\exp\{-\sum_{a,s,s'}k_\mu^{(a)} (s)
D_0(s-s')k_\mu^{(a)} (s')\nonumber\\ 
&&+2\pi i \sum_{s}N_\mu(s)k_\mu^{(1)}(s)
+i \sum_{a,s}B^{(a)} (s){\partial_\mu}'k_\mu^{(a)} (s)
+i\sum_{s}\phi_\mu (s)\sum_a k_\mu^{(a)} (s)\nonumber\\
&&+i\sum_{a,s^{(n)}}\gamma_\mu^{(a)} (s^{(n)})(K_\mu^{(a)} (s^{(n)})
-\sum_{i,j,k=0}^{n-1} k_\mu^{(a)})\} 
\end{eqnarray}
where we have introduced auxiliary fields $\phi$,\ $B^{(a)}$
and $\gamma^{(a)}$ for the constraints $\sum_{a} k^{(a)} =0$,\ 
${\partial'}_\mu k_\mu^{(a)} (s)=0$ and the definition of
$K^{(a)}$,\ respectively.\ Here we have used such notations as
${\cal D}\mbox{\boldmath $\gamma$}\equiv \prod_a {\cal D}\gamma^{(a)}$.\
There are following identities for integers
$m^{(a)} (s),\ n_\mu^{(a)} (s^{(n)})$ and $P_\mu (s)$: 
\begin{eqnarray}
&&\exp\{2\pi i m^{(a)} (s){\partial_\mu}'k_\mu^{(a)}(s)\}=1,\\
&&\exp\{2\pi i n_\mu^{(a)} (s^{(n)})(K_\mu^{(a)} (s^{(n)})-
\sum_{i,j,k=0}^{n-1}k_\mu^{(a)})\}=1,\\
&&\exp\{2\pi i P_\mu (s)\sum_a k_\mu^{(a)} (s)\}=1.
\end{eqnarray}
Then we can change the integral region of
$\gamma^{(a)},\ B^{(a)}$ and $\phi$
from the first Brillouin zone to the infinite region.\
Using the Poisson summation formula
\begin{eqnarray}
\sum_{k_\mu^{(a)} \in Z }f[k_\mu^{(a)}]&=&
{\rm const}.\int_{-\infty}^{\infty}
dF_\mu^{(a)} \sum_{l_\mu^{(a)} \in Z}
\exp \{2\pi i F_\mu^{(a)} l_\mu^{(a)} \}f[F_\mu^{(a)}],
\end{eqnarray}
$Z[K^{(a)},J]$ becomes
\begin{eqnarray}
Z[K^{(a)},J]&=&\int_{-\infty}^{\infty}{\cal D}\mbox{\boldmath
$\gamma$}
\int_{-\infty}^{\infty}
{\cal D}\phi 
\int_{-\infty}^{\infty}{\cal D}\mbox{\boldmath
$B$}\int_{-\infty}^{\infty}
{\cal D}\mbox{\boldmath $F$}
\sum_{l^{(a)} \in Z}\nonumber\\
&&\times\exp\{-\sum_{a,s,s'}F_\mu^{(a)}
(s)D_0(s-s')F_\mu^{(a)} (s')
+2\pi i \sum_{s}N_\mu (s)F_\mu^{(1)} (s) \nonumber\\
&&+ i\sum_{a,s}(-\partial_\mu B^{(a)}
(s)
+\phi_\mu (s)
+2\pi l_\mu^{(a)} (s))F_\mu^{(a)} (s)\nonumber\\
&&-i\sum_{a,s^{(n)}}\gamma_\mu^{(a)} (s^{(n)})\sum_{i,j,k=0}^{n-1}
F_\mu^{(a)} 
\}.
\end{eqnarray}
Writing the lattice spacing explicitly,\ we get
\begin{eqnarray}
&&-ia^4 n\sum_{a,s^{(n)}}\gamma_\mu^{(a)} (nas^{(n)})\sum_{i,j,k=0}^{n-1}
F_\mu^{(a)} (nas^{(n)}+(n-1)a\hat\mu+ia\hat\nu+ja\hat\rho
+ka\hat\sigma)\nonumber\\
&&=-ia^4\sum_{a,s}X_\mu^{(a)} (as)F_\mu^{(a)} (as),
\end{eqnarray}
\begin{eqnarray}
X_\mu^{(a)}(as)&\equiv&na^4\sum_{s^{(n)}}\gamma_\mu^{(a)}
(nas^{(n)})\nonumber\\
&&\times \delta(nas_\mu^{(n)}+(n-1)a-as_\mu)
\prod_{i(\ne\mu)}\sum_{I=0}^{n-1}\delta(nas_i^{(n)}+Ia-as_i).
\end{eqnarray}
Integrating out $F^{(a)}$ 
and $B^{(a)}$,\ we find
\begin{eqnarray}
Z[K^{(a)},J]&=&
\int_{-\infty}^{\infty}{\cal D}\mbox{\boldmath $\gamma$}
\exp\{i b^4\sum_{a,s^{(n)}}\gamma_\mu^{(a)} (b s^{(n)})K_\mu^{(a)} (b s^{(n)})
\}\int_{-\infty}^{\infty}{\cal D}\phi\nonumber\\
&&\times
\exp\{-\frac14 a^8 \sum_{s,s'}V_\mu^{(a)}
(as)A_{\mu\nu}(as-as')V_\nu^{(a)}(as')\}
\nonumber\\
&&
\times
\sum_{al^{(a)} \in Z}\exp\{-\pi^2 a^8 \sum_{a,s,s'}l_\mu^{(a)}(as)A_{\mu\nu}
(as-as')l_\nu^{(a)}(as')\nonumber\\
&&+\pi^2 a^8 \sum_{a,s,s'}V_\mu^{(a)} (as)A_{\mu\nu}
(as-as')l_\nu^{(a)} (as)\},
\end{eqnarray}
where
\begin{eqnarray}
&&V_\mu^{(a)}(as)\equiv X_\mu^{(a)}(as)-\phi_\mu (as)-2\pi N_\mu (as)
\delta_{a,1},\\
&&A_{\mu\nu}(as-as')\equiv
  \Biggl\{
    \delta_{\mu\nu}-\frac{\partial_\mu\partial'_\nu}{\sum_{\rho}\partial_\rho
     \partial'_\rho}
  \Biggr\}
  D_0^{-1}(as-as').
\end{eqnarray}
Here we consider $l^{(a)}=0$ alone,\ since we finally take the $a\to 0$ limit
in the sum with respect to $a l^{(a)} \in Z$.\
We change the variables from $\gamma^{(a)}$ to ${\gamma'}^{(a)}$:
\begin{eqnarray}
{\gamma'}^{(1)}&\equiv& \gamma^{(2)}-\gamma^{(3)},\
{\gamma'}^{(2)}\equiv \gamma^{(3)}-\gamma^{(1)},\
{\gamma'}^{(3)}\equiv \gamma^{(1)}-\gamma^{(2)}.
\end{eqnarray}
Integrating  out ${\gamma'}^{(a)}$ and $\phi_\mu$,\ we get the monopole action in
terms of $K^{(a)}$ as follows:
\begin{eqnarray}
Z[J]&=&\exp\{-\frac23 \pi^2
a^8\sum_{s,s'}N_\mu(as)A_{\mu\nu}(as-as')N_\nu(as')
\nonumber\\
&&+\frac23 \pi^2 b^8
\sum_{s^{(n)},{s'^{(n)}}}B_\mu(bs^{(n)}){A'}_{\mu\nu}^{-1}(
bs^{(n)}-b{s'^{(n)}})B_\nu(b{s'^{(n)}})\}Z_{mon}[J],
\end{eqnarray}
where
\begin{eqnarray}
&&B_\mu(bs^{(n)})\equiv\lim_{a\to 0 \atop{n\to\infty}}
  \frac{1}{n^3}a^8\sum_{s,s',\nu}
    \delta ( nas_\mu^{(n)}+(n-1)a-as_\mu )
\prod_{i(\ne \mu)}
      \sum_{I=0}^{n-1}\delta (nas_i^{(n)}+Ia-as_i )\nonumber\\
&&\quad\quad\quad \times
  \left\{
    \delta_{\mu\nu}
    -\frac{\partial_{\mu}\partial'_{\nu}}{\sum_{\rho}\partial_{\rho}
           \partial'_{\rho}}
  \right\}
D_0^{-1}(as-as')N_{\nu}(as'),\\
&&{A'}_{\mu\nu}(bs^{(n)}-bs'^{(n)})
\equiv
\nonumber\\
&&\qquad
\lim_{a\to 0 \atop{n\to\infty}}
\frac{1}{n^6}a^8\sum_{s,s'}
  \delta ( nas_\mu^{(n)}+(n-1)a-as_\mu )
  \prod_{i(\ne \mu)}
    \sum_{I=0}^{n-1}\delta (nas_i^{(n)}+Ia-as_i )\nonumber\\
&&\qquad\qquad\;\times
  \left\{
    \delta_{\mu\nu}
    -\frac{\partial_{\mu}\partial'_{\nu}}{\sum_{\rho}\partial_{\rho}
     \partial'_{\rho}}
  \right\}
  D_0^{-1}(as-as')
\nonumber\\
&&\qquad\qquad\;\times
  \delta ( na{s'}_\nu^{(n)}+(n-1)a-as_\nu' )
  \prod_{j(\ne \nu)}
    \sum_{J=0}^{n-1}\delta ( na{s'}_j^{(n)}+Ja-as_j' ).
\end{eqnarray}
$Z_{mon}[J]$ is the dynamical monopole part:
\begin{eqnarray}
Z_{mon}[J]&=& \sum_{K^{(a)} \in Z}
\delta_{\partial'_\mu
K_\mu^{(a)} ,0}\delta_{\Sigma_a K^{(a)},0}\nonumber\\
&&\times\exp\{
  - b^8 \sum_{a,s^{(n)},s'^{(n)}}
K_{\mu}^{(a)}(bs^{(n)})
  A_{\mu\nu}^{\prime -1}(bs^{(n)}-bs'^{(n)})
K_{\nu}^{(a)}(bs'^{(n)})\nonumber\\
&&
+  2 \pi i b^8 \sum_{s^{(n)},s'^{(n)}}
  B_{\mu}(bs^{(n)})A_{\mu\nu}^{\prime -1}(bs^{(n)}-bs'^{(n)})
  K_{\nu}^{(1)}(bs'^{(n)})\}.
\end{eqnarray}
The interactions of the perfect action on the $b$-lattice 
in Eq.\ (4.20) depend on directions.\
This is consistent with the numerical data.\par
The spectrum of $K_\mu^{(a)}(bs^{(n)})$ is found to be equivalent to
that in the continuum theory as discussed in the SU(2) case [26].\

\section{String representation and rotational invariance}

When we transform $Z_{mon}[J]$ in Eq.\ 
(4.20) into the string representation as in SU(2) [31,32],\
we can estimate the static potential analytically.\par
First we change the variables as follows:
\begin{eqnarray}
K^{(1)}&=&j^{(2)}-j^{(3)},\ K^{(2)}=j^{(3)}-j^{(1)},\ K^{(3)}
=j^{(1)}-j^{(2)}.
\end{eqnarray}
The conservation laws of monopole currents $\partial'_\mu K_\mu^{(a)}=0$
leads us to
$\partial'_\mu j^{(1)}_\mu=\partial'_\mu j^{(2)}_\mu 
=\partial'_\mu j^{(3)}_\mu$.\
The conditions are expressed as 
\begin{eqnarray}
&&\int_{-\pi}^{\pi}{\cal D}\mbox{\boldmath $\varphi$}
\delta(\sum_a \varphi^{(a)})
\exp\{i\sum_{a,s} j_\mu^{(a)}(s)\partial_\mu\varphi^{(a)}(s)\}.
\end{eqnarray}
Then $Z_{mon}[J]$ is reduced to the following:
\begin{eqnarray}
Z_{mon}[J]&=&
\int_{-\pi}^{\pi}{\cal D}\mbox{\boldmath $\varphi$}
\delta(\sum_a \varphi^{(a)})\sum_{j^{(a)} \in Z}
\exp\{-2\sum_{a,s,s'}j_\mu^{(a)}(s){A'}_{\mu\nu}^{-1}(s-s')
j_\nu^{(a)}(s')\nonumber\\
&&+2\sum_{a < b ,s,s'}
j_\mu^{(a)}(s){A'}_{\mu\nu}^{-1}(s-s')j_\nu^{(b)}(s')
+i\sum_{a,s}j_\mu^{(a)}(s)\partial_\mu\varphi^{(a)}(s)\nonumber\\
&&+2\pi i \sum_{s,s'}B_\mu(s){A'}_{\mu\nu}^{-1}(s-s')(j_\nu^{(2)}(s')-
j_\mu^{(3)}(s'))\}\\
&=&\int_{-\infty}^{\infty}{\cal D}{\overrightarrow C}
\int_{-\pi}^{\pi}{\cal D}{\mbox{\boldmath
$\varphi$}}\delta(\sum_a
\varphi^{(a)})\sum_{j^{(a)} \in Z}\exp\{-\frac18 \sum_{s,s'}
{\overrightarrow C}_\mu(s){A'}_{\mu\nu}(s-s'){\overrightarrow C}_\nu(s')
\nonumber\\
&&+i\sum_{a,s} j_\mu^{(a)}(s)(\partial_\mu
\varphi^{(a)}(s)+{\vec \epsilon}_a
\cdot \overrightarrow C_\mu (s))\nonumber\\
&&+2\pi i \sum_{s,s'}B_\mu(s){A'}_{\mu\nu}^{-1}(s-s')
(j_\nu^{(2)}(s')-j_\nu^{(3)}(s'))\},
\end{eqnarray}
where auxiliary fields $\overrightarrow{C}_{\mu} 
\equiv (C_{\mu}^3, C_{\mu}^8)$ are
introduced and  
$\vec{\epsilon}_{a}$
are the SU(3) root vectors:
$\vec{\epsilon}_1=(1,0)$,\  $\vec{\epsilon}_2=(-1/2,-\sqrt{3}/2)$,\
$\vec{\epsilon}_3=(-1/2,\sqrt{3}/2)$.\
Since $\sum_a \varphi^{(a)}=0$ and $\sum_a {\vec \epsilon}_a =0$,\
we get an identity 
\begin{eqnarray}
\exp\{-\frac{i}{2\pi}\sum_{s}\phi_\mu(s)
\sum_a (\partial_\mu\varphi^{(a)}(s)+{\vec \epsilon}_a \cdot
{\overrightarrow C}_\mu(s))\}=1.
\end{eqnarray}
The Poisson summation formula
\begin{eqnarray}
\sum_{l_\mu^{(a)} \in Z}\exp\{2\pi i 
\sum_{a}F_\mu^{(a)}l_\mu^{(a)}
+i \phi_\mu \sum_a l_\mu^{(a)}\}&=&
\sum_{j_\mu^{(a)} \in Z}\prod_a \delta
 (F_\mu^{(a)}+\frac{\phi_\mu}{2\pi}-j_\mu^{(a)})
\end{eqnarray}
gives us 
\begin{eqnarray}
&&\int_{-\infty}^{\infty}{\cal D\mbox{\boldmath $F$}}\sum_{l^{(a)}\in
Z}\delta_{\Sigma_a l^{(a)} ,0}
\exp\{2\pi i\sum_{a,s}F_\mu^{(a)}(s)l_\mu^{(a)}(s)\}f(
F_\mu^{(a)})=\nonumber\\
&&\ \ \ \ \ \ \ \ \ \ \ \ \ \ \ \ \ \ \ \ \ \ \ \ \ \ \ \ \ \ \ \ \ \
\ \ \ \ \ \ 
{\rm {const}}.\int_{-\infty}^{\infty}{\cal D}\phi\sum_{j^{(a)} \in Z}
f(j_\mu^{(a)} -
\frac{\phi_\mu}{2\pi}).
\end{eqnarray}
Then $Z_{mon}[J]$ becomes
\begin{eqnarray}
Z_{mon}[J]&=&\int_{-\infty}^{\infty}{\cal D}\mbox{\boldmath $F$}
\int_{-\infty}^{\infty}{\cal D}\overrightarrow C
\int_{-\pi}^{\pi}{\cal D}\mbox{\boldmath $\varphi$}\delta(\sum_a
\varphi^{(a)})\sum_{l^{(a)} \in Z}\delta_{\Sigma_a l^{(a)} ,0}\nonumber\\
&&\times\exp\{-\frac18 \sum_{s,s'}
{\overrightarrow C}_\mu(s){A'}_{\mu\nu}(s-s')
{\overrightarrow C}_\nu(s')\nonumber\\
&&+i\sum_{a,s}F_\mu^{(a)}(s)
(\partial_\mu\varphi^{(a)}(s)+{\vec \epsilon}_a
\cdot{\overrightarrow C}_\mu(s) +2\pi l_\mu^{(a)}(s))\nonumber\\
&&
+2\pi \sum_{s'}(F_\mu^{(2)}(s)-F_\mu^{(3)}(s))
{A'}_{\mu\nu}^{-1}(s-s')B_\nu(s')\}.
\end{eqnarray}
Here we perform the Berezinski-Kosterlitz-Thouless (BKT)
transformation [33]:
\begin{eqnarray}
&&l_\nu^{(a)}(s)=s_\mu^{(a)}(s) + \partial_\mu r^{(a)}(s)
,\ \partial_{[\mu}s_{\nu]}^{(a)}(s)=\sigma_{\mu\nu}^{(a)}(s),\
\nonumber\\
&&\partial_\mu\varphi^{(a)}(s) +2\pi l_\mu^{(a)}(s)
=\partial_\mu\varphi_{nc}^{(a)}(s)
-2\pi\sum_{s'}\partial'_\nu\Delta^{-1} (s-s')
\sigma_{\nu\mu}^{(a)}(s'),\
\end{eqnarray}
where
$\varphi_{nc}^{(a)}(s)
\equiv \varphi^{(a)}(s)-2\pi\sum_{s'}\Delta^{-1} (s-s')\partial'_\nu 
s_\nu^{(a)}(s')
+2\pi r^{(a)}(s)$ is non-compact.\
The plaquette variable $\sigma_{\mu\nu}^{(a)}$ satisfies a conservation
law $\partial{}_{[\alpha}\sigma_{\mu\nu]}^{(a)}(s)=0$ 
and a constraint 
$\sum_a \sigma_{\mu\nu}^{(a)} =0$ due to 
$\sum_a l_{\mu}^{(a)} =0$.\ 
Using the condition $\partial'_\mu B_\mu=0$,\ we integrate out
$F_\mu^{(a)}$,\
${\overrightarrow C}_\mu$ and $\varphi_{nc}^{(a)}$.\ We get the string
representation:
\begin{eqnarray}
Z_{str}[J]&=&
\exp\{-\frac23 \pi^2
\sum_{s,s'}
B_\mu(s){A'}_{\mu\nu}^{-1}(
s-s')B_\nu(s')\}
\sum_{\sigma^{(a)} \in Z}
\delta_{\partial_{[\alpha}\sigma_{\mu\nu]}^{(a)},0}
\delta_{\Sigma_a \sigma^{(a)},0}\nonumber\\
&&\times\exp\{
-\frac{\pi^2}{3}\sum_{a,s,s'}
\partial'_\alpha \sigma_{\mu\alpha}^{(a)} (s)H_{\mu\nu}(s-s')
\partial'_\beta \sigma_{\nu\beta}^{(a)} (s')\nonumber\\
&&+\frac{2 \pi^2}{3}\sum_{s,s'}
(\sigma_{\mu\nu}^{(2)}(s)-\sigma_{\mu\nu}^{(3)}(s))\partial_\nu
\Delta^{-1}(s-s')B_\mu(s')\},
\end{eqnarray}
\begin{equation}
H_{\mu\nu}(s-s')=\sum_{s_1}{A'}_{\mu\nu}(s-s_1)\Delta^{-2}(s_1 -s').
\end{equation}
\par
Since the numerical results show that the couplings of the monopole action are weak
for large $b$ region,\ we can use the strong coupling expansion
in the dual string model.\ 
Then we see that the quantum fluctuations can be neglected as in SU(2) 
case [26].\ 
The classical part in Eq.\ (5.10)
cancels the second classical term in Eq.\ (4.17).\
\ As a result,\ 
the classical part of the expectation value of the
Wilson loop is reduced to
\begin{eqnarray}
\langle W(C) \rangle_{cl}&=&
\exp\{-\frac23 \pi^2
\int_{-\infty}^{\infty}d^4 x d^4 y
N_\mu(x)A_{\mu\nu}(x-y)N_\nu(y)\}.
\end{eqnarray}
Since the classical part is written in the continuum form,\ 
the continuum rotational invariance is trivial.\ 
The static potential $V(Ib,0,0)$ and $V(Ib,Ib,0)$ can be
evaluated as in SU(2).\
We take the following plaquette sources as $S_{\alpha\beta}(x)$
for $V(Ib,0,0)$:
\begin{eqnarray}
S_{\alpha < \beta}(x)&=&
\delta_{\alpha 1}\delta_{\beta 4}
\theta (x_1)\theta (Ib-x_1)\theta (x_4)\theta (Tb-x_4)
\delta (x_2)\delta (x_3)
\end{eqnarray}
and for $V(Ib,Ib,0)$:
\begin{eqnarray}
S_{\alpha < \beta}(x)&=&(\delta_{\alpha 1}\delta_{\beta 4}
+\delta_{\alpha 2}\delta_{\beta 4})\delta (x_3)\theta (x_4)
\theta (Tb-x_4)\nonumber\\
&&\ \ \ \ \ \ \ \ \ \ \ \ \ \ \ \ \ \ \ 
\times\theta (x_1)\theta (Ib-x_1)\theta (x_2)
\theta (Ib-x_2)\delta (x_1-x_2),
\end{eqnarray}
respectively.\ 
Using the following formula
\begin{eqnarray}
\lim_{T\to \infty}\left(\frac{\sin \alpha T}{\alpha}\right)^2
&=&\pi T \delta (\alpha),
\end{eqnarray}
we get the static potentials
\begin{eqnarray}
V(Ib,0,0)&=&-\lim_{T\to\infty}\frac1{Tb}\ln \langle W(
Ib,0,0,Tb) \rangle_{cl}\nonumber\\ 
&\mapright{I\to\infty}&\frac{2\pi^2}{3}(Ib)^2\int \frac{d^2 p}{2\pi^2}
\left[ \frac1{\Delta D_0}\right](0,p_2,p_3,0)\nonumber\\
&=&\frac{\pi\kappa Ib}{3}\ln\frac{m_1}{m_2},\\
V(Ib,Ib,0)&=&\frac{\sqrt{2}\pi\kappa Ib}{3}\ln\frac{m_1}{m_2},
\end{eqnarray}
where
$\kappa (m_1^2-m_2^2)=\gamma^{-1}, m_1^2+m_2^2=\beta/\gamma 
,m_1^2m_2^2=\alpha/\gamma$.
The static potential has the linear term alone and the rotational
invariance is recovered completely.\ The string tension is evaluated as
\begin{equation}
\sigma_{cl}=\frac{\pi\kappa}{3}\ln\frac{m_1}{m_2}.
\end{equation}
This is consistent with the results [34].\ 
The $m_1^{-1}$ and the $m_2^{-1}$ could be
regarded as the coherence and the penetration lengths in Type-2
superconductor.\

\section{Estimate of the string tension}
\begin{figure}[tb]
\begin{minipage}{73mm}
\epsfxsize=7.3cm
\begin{center}
\leavevmode
\epsfbox{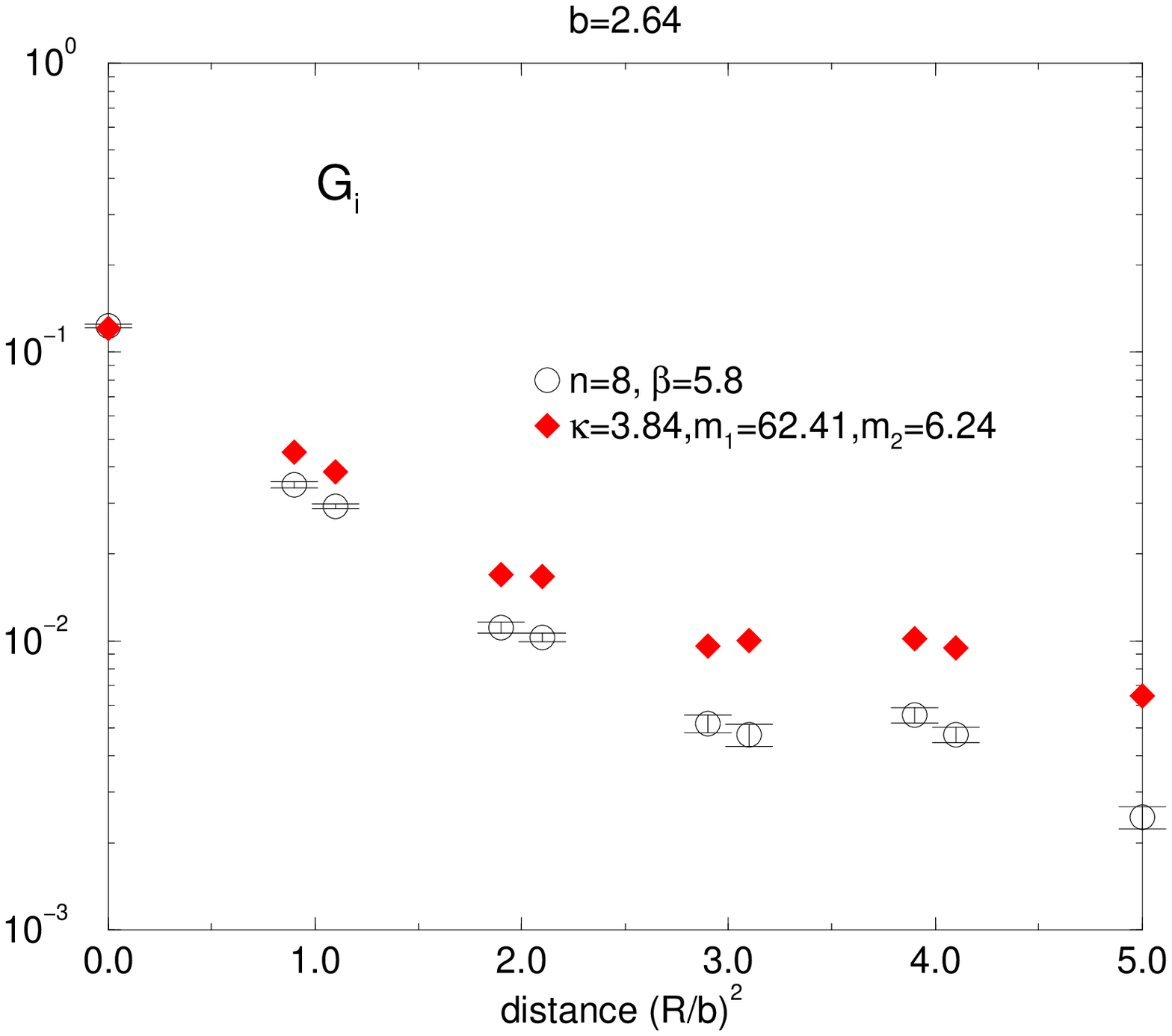}
\end{center}
\label{fig:9}
\end{minipage}
\hfill
\begin{minipage}{73mm}
\epsfxsize=7.3cm
\begin{center}
\leavevmode
\epsfbox{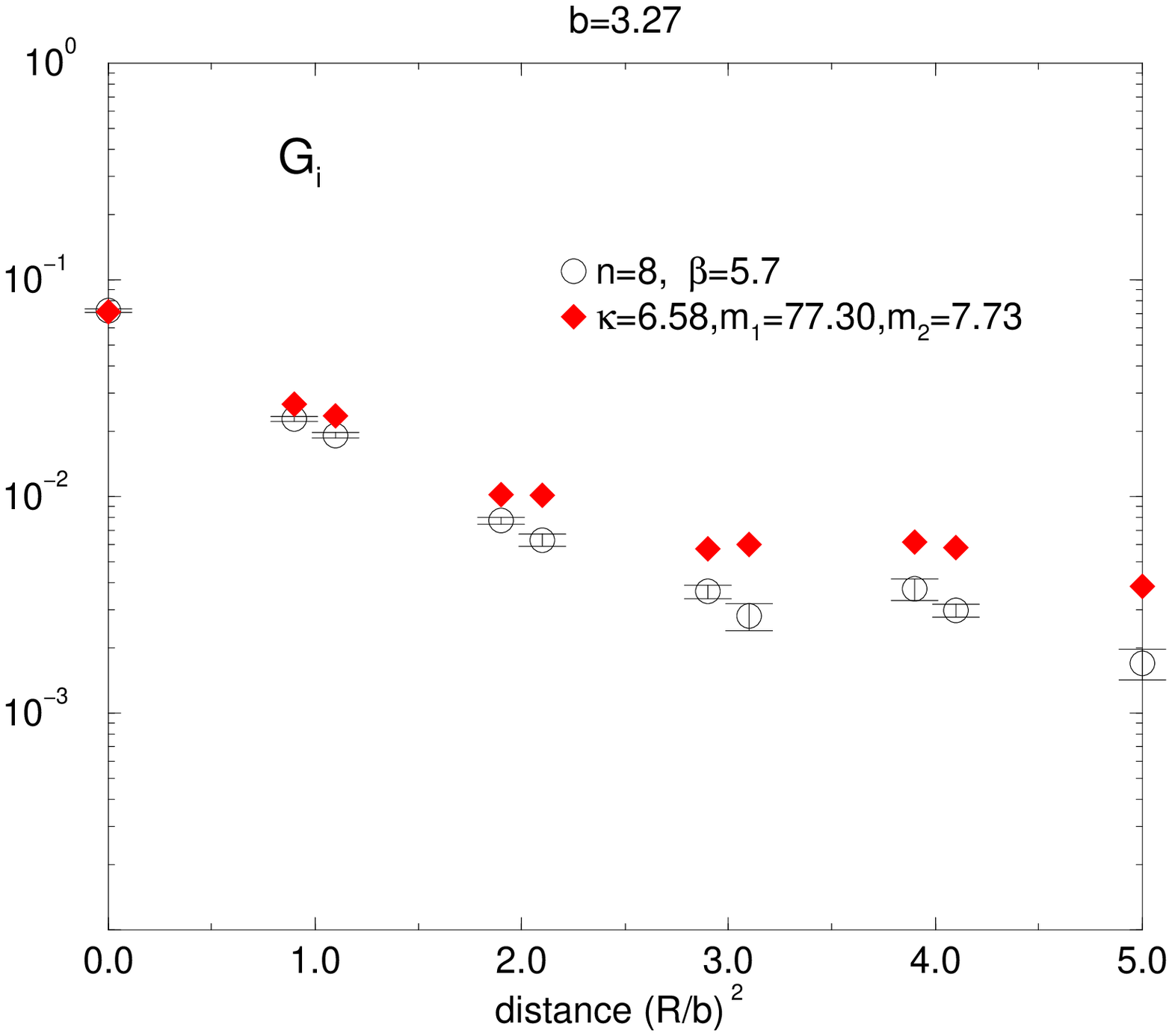}
\end{center}
\label{fig:9}
\end{minipage}

\epsfxsize=7.3cm
\begin{center}
\leavevmode
\epsfbox{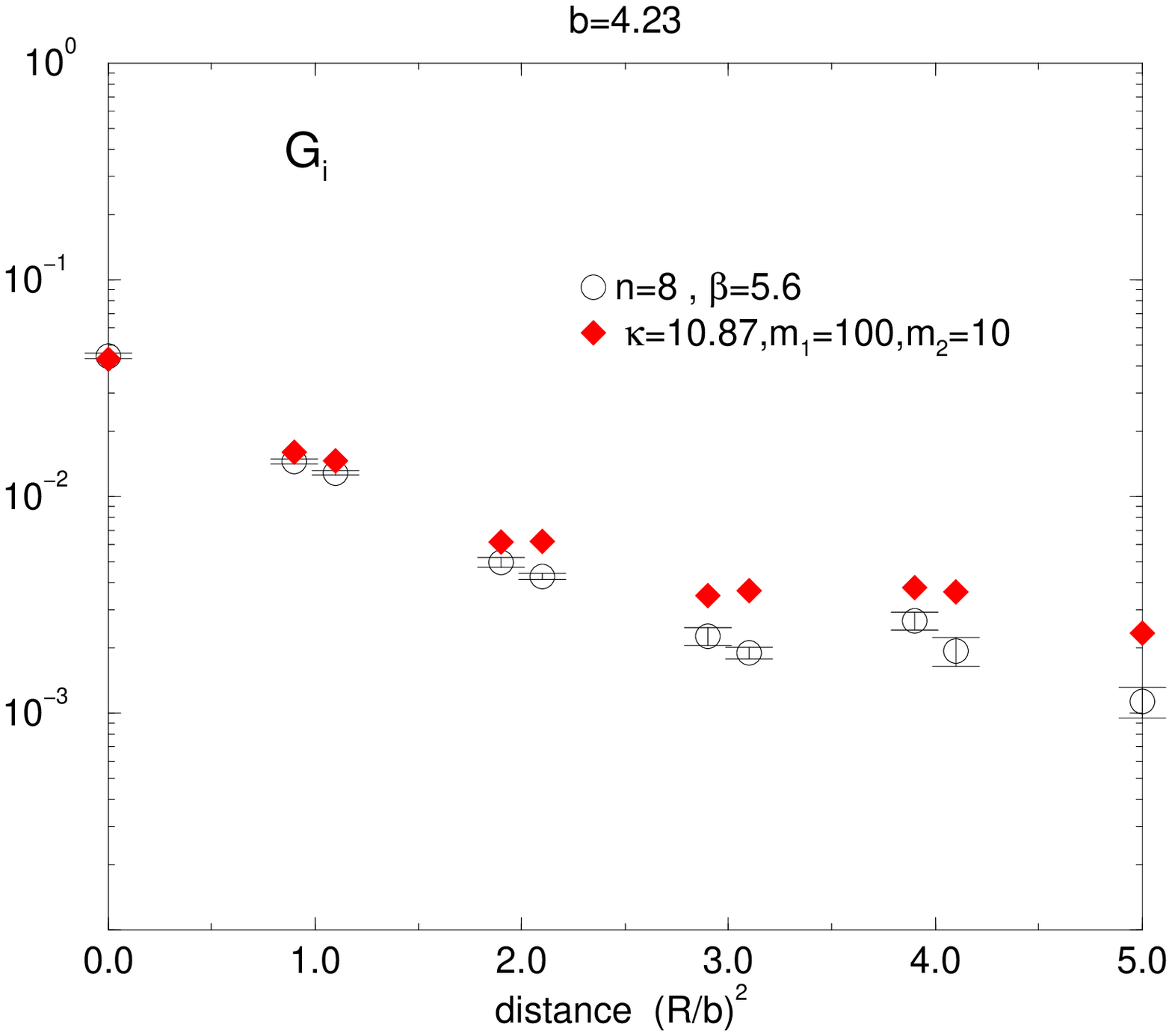}
\end{center}
\caption{ 
The coupling $G_i$ versus distance.\
The data from (i) perfect action (diamonds)
and (ii) Swendsen method (circles).\
The values of dimensionless $\kappa$,\ $m_1$ and $m_2$ are shown. }
\label{fig:9}
\end{figure}

Ths string tension in Eq.\ (5.18) is written by the parameters
of the monopole action taken
on the $a$-lattice.\
The parameters can be determined by comparing 
the theoretical perfect action in Eq.\ (4.20) with
the numerical results
of the monopole action on the $b$-lattice.\ 
${A'}_{\mu\nu}^{-1}$ in Eq.\ (4.19) fixed the gauge is calculated in 
ref.\ [26].\ 
The results of the comparison are
seen in Figure 3.\
The agreement is rather good.\ 
The value of the string tension is 
$\sqrt{\sigma_{cl}/\sigma_{ph}}\sim 1.22$ for large $b$
as in Figure 4.\
The result is not so bad,\ although the discrepancy is not negligible.\par
 In order to study whether the difference is due to the 
ambiguity of the fit of $D_0$,\
we try to estimate the string tension without
determining the parameters on the $a$-lattice.\ 
The following monopole Wilson loop on the $b$-lattice is considered:
\begin{eqnarray}
\langle W(C)\rangle&=&\frac1{Z} \sum_{K^{(a)} \in Z}\delta_{\partial'_\mu
K_\mu^{(a)} ,0}\delta_{\Sigma_a K^{(a)},0}\exp\{
- \sum_{a,s,s'}
K_{\mu}^{(a)}(s)
  D_{\mu\nu}^{-1}(s-s')
K_{\nu}^{(a)}(s')\nonumber\\
&&
+2\pi i\sum_{s}N_\mu(s)K_\mu^1(s)\},
\end{eqnarray}
where the naive monopole Wilson loop operator (A.15) on the coarse $b$-lattice
is used
and 
$D_{\mu\nu}^{-1}$ corresponds to ${A'}_{\mu\nu}^{-1}$.\ 
Transforming Eq.\ (6.1) to the string representation and neglecting the 
quantum fluctuations as in the previous section,\ we get the classical part 
of the Wilson loop as follows:
\begin{equation}
\langle W(C)\rangle_{cl}=\exp\{-\frac23 \pi^2
\sum_{s,s'}N_\mu(s)D_{\mu\nu}
(s-s')N_\nu(s')\}.
\end{equation}
It is found theoretically that 
the static potential $V(Ib,0,0)$ evaluated by Eq.\ (6.2) agrees with 
Eq.\ (5.16) in the $I\to\infty$ case.\ 
Using Eq.\ (6.2),\ the string tension becomes 
\begin{eqnarray}
\sigma_{cl}&=&
\frac{\pi^2}6 \int_{-\pi}^{\pi}
\frac{d^2 k}{(2\pi)^2}\frac1{({\sin}^2 \frac{k_2}{2}
+{\sin}^2 \frac{k_3}{2})^2}\nonumber\\
&&\times
[{\sin}^2\frac{k_3}{2}D(0,k_2,k_3,0;\hat2)
+{\sin}^2\frac{k_2}{2}D(0,k_2,k_3,0;\hat3)],
\end{eqnarray}
\begin{equation}
D_{\mu\nu}(s-s')
=\delta_{\mu\nu}\int_{-\pi}^{\pi}\frac{d k^4}{(2\pi)^4}
D(k_1,k_2,k_3,k_4;\hat{\mu})e^{ik(s-s')}.
\end{equation}
We estimate the string tension from
Eq.\ (6.3) with the numerical data of the monopole action on the
$b$-lattice.\par 
The results   
are seen in Figure 4.\
The results are not so different from those of the first method
for large $b$.\ The $b$-independence holds roughly for $b > 2$ and
$\sqrt{\sigma_{cl}/\sigma_{ph}}\sim 1.27$ for large $b$.\ 
The difference from the physical value is also seen in this
method.\par 
Using the same method as in SU(2) [26],\ the effect of the 
quantum fluctuation can be evaluated.\
The order of the correction for the string tension becomes
\begin{equation}
-\frac{8}{b^2}\exp \{-5 \Pi (0) b^2 \},
\end{equation}
where $\Pi (0)$ represents the coupling of
the self-interaction in the string model.\
For example,\ 
the value of Eq.\ (6.5) is $-2.6\times 10^{-10}$ for $b=3.27$.\
The correction is very small in large $b$ region.\par
In order to clarify the origin of the difference,\ 
we have to study the effects of systematic errors 
carefully.\ 

\begin{figure}[tb]
\epsfxsize=10cm
\begin{center}
\leavevmode
\epsfbox{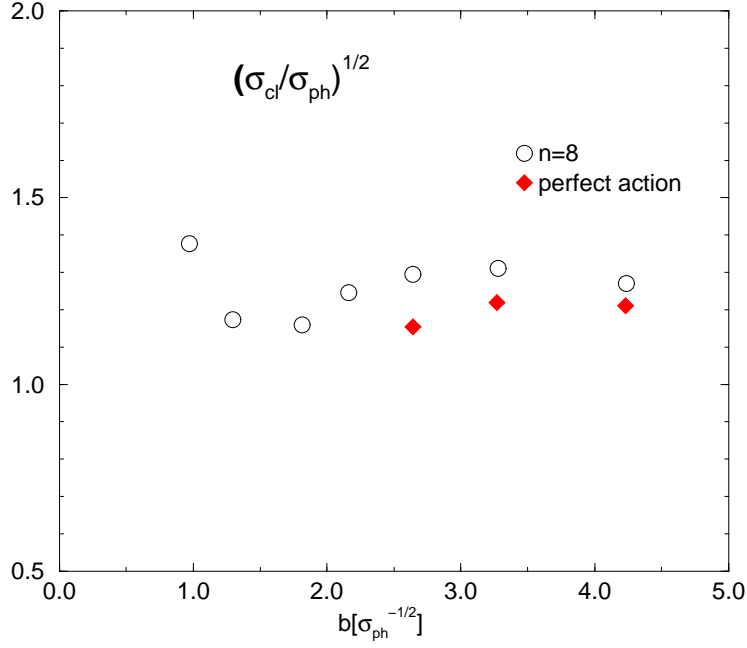}
\end{center}
\caption{ 
 The string tension versus b.\ 
The data from (i) Eq.\ (5.18) (diamonds) and (ii) Eq.\ (6.3) (circles).}
\label{fig:9}
\end{figure}

As a systematic error,\ the difference between D-T monopoles used in the
simulation and the monopoles in the analytical calculations
may be important.\
The magnetic charge of the latter ranges from $-\infty$ to $\infty$,\
while D-T monopoles take only the restricted values of magnetic
charges.\ This is under investigation.\par
The glueball spectrum can be evaluated similarly as in SU(2) QCD [26].\
The lightest $0^{++}$ glueball mass is determined as $M_{0^{++}} = 2 m_2$ .\
For $b=4.23$,\ $2 m_2/\sqrt{\sigma_{ph}}= 3.90$.\ 
The result is not so different from $3.64\pm 0.09$ [35].\

\section{Conclusions}
We have studied an effective monopole action in pure SU(3) lattice
QCD.\
The following results are obtained:
\begin{enumerate}
\item
We have found that the SU(3) monopole action can be derived clearly
for various $n$-blocked monopoles in MA gauge numerically,\
using the extended Swendsen method in the two-current and the one-current
cases.\
We perform $n=2\sim8$ step block-spin transformations on the $48^4$
lattice.\
In the two-current case,\ 
we obtain an almost perfect lattice monopole action 
for the infrared region of SU(3) QCD,\
since the action seems to depend on a physical scale $b$ alone.\
The simple 4,\ 6-point interactions become negligibly small as compared
with two-point interactions for large $b$ region as in SU(2).\
Thus two-point monopole interactions are dominant in the
infrared region.\
In the one-current case,\
if the physical scale $b$ is taken in unit of the string tension 
$\sqrt{\sigma_{ph}}$,\
the scaling which was not seen in the previous study [27]
holds beautifully.\
Monopole condensation occurs in the large $b$ region
from the energy-entropy balance.\
\item
The perfect action satisfies the continuum rotational
invariance.\
When we try to study the restoration of the rotational invariance
of the static potential,\
we have to know the correct
form of the perfect operator.\
For that purpose,\
a block-spin transformation
from the small $a$-lattice ($a\to0$) to the finite
$b=na(\beta)$ lattice is performed analytically in a simple restricted 
case of a quadratic monopole action as in SU(2) case.\
The perfect monopole action and the perfect operator evaluating the
static potential between electric charges on the $b$-lattice are
derived.\
The quadratic SU(3) monopole action is simple,\ but it corresponds to
the London limit of the DGL theory which is non-trivially interacting 
field theory.\

\item
The SU(3) monopole action can be transformed into the string model.\
There are three strings $\sigma^{(a)}$ satisfying one constraint
$\sum_a \sigma^{(a)} =0$.\
This is consistent with the results in the continuum limit [36,37].\
Since the monopole interactions are weak for large $b$ region,\
the dual string interactions are strong.\
Using the strong coupling expansion,\
the static potential in the long distance is calculated by the
classical part alone analytically.\
The static potential has a linear term alone and
the restoration of the rotational invariance can be seen explicitly.\
The string tension is estimated from the numerical data of the monopole
action.\
The result is rather good,\ but the difference from the physical string
tension is not negligible.\
The same thing happens in SU(2) case [26].\ 
It seems that there still exist some systematic errors.\
\end{enumerate}

\section*{Acknowledgments}
This work is supported by the Supercomputer Project (No.98-33
and No.99-47) of
High Energy Accelerator Research Organization (KEK) and the
Supercomputer Project of the Institute of Physical and Chemical
Research (RIKEN).\ T.S.\ is financially supported by JSPS Grant-in-Aid
for Scientific Research (B) (No.\ 10440073 and No.\ 11695029).\

\appendix

\section{Appendix}

\setcounter{equation}{0}
\renewcommand{\theequation}
{A.\arabic{equation}}
In SU(2),\ it is already known that the 
monopole action can be derived from the dual
abelian higgs model on the fine $a$-lattice [8,32,38].\
We show how to derive the monopole action from the DGL theory
on the lattice in SU(3).\
In DGL,\ $\overrightarrow{A}_{\mu} \equiv (A_{\mu}^3 , A_{\mu}^8
) $ are used as $U(1)^2$ photon fields,\ where
$A_{\mu}=A_{\mu}^i\lambda^i /2$\ ($\lambda^i$ are Gell-Mann matrices).\
The magnetic charges $\overrightarrow{m}$ are distributed on the SU(3) 
root lattice:
$\overrightarrow{m} = g_m\sum_{a=1}^3
\xi_{a}\vec{\epsilon}_{a}$,\
$\xi_{a} \in Z$,\ $g_m=4\pi/g$ ($g$ is the SU(3) coupling constant).\ 
The DGL Lagrangian in the continuum limit [34] is 
\begin{equation}
{\cal
L}_{DGL}=\frac14(\partial_{\mu}\overrightarrow{C}_{\mu}-\partial_{\nu}
\overrightarrow{C}_{\mu})^2 +
\sum_{a=1}^3(|(\partial_{\mu}+ig_m\vec{\epsilon}_{a}\cdot\overrightarrow
{C}_{\mu})\phi^{(a)}|^2 + \lambda(|\phi^{(a)}|^2 -  v^2)^2),
\end{equation}
where $\overrightarrow{C}_{\mu} \equiv (C_{\mu}^3 , C_{\mu}^8)$ are
dual $U(1)^2$  gauge fields and 
$\phi^{(a)}=\rho^{(a)} \exp(i\varphi^{(a)})$ are monopole fields with a
constraint $\sum_a \varphi^{(a)}=0$.\par 
In the below,\ we use the notations of differential forms on the
lattice [39].\
The lattice DGL action [40] becomes
\begin{eqnarray}
S_{DGL}[\overrightarrow{C},\phi^{(a)}]&=&\frac{1}{2g_m^2}
||d\overrightarrow{C}||^2
-\gamma\sum_{a,s,\mu}(\phi_s^{(a)\ast}U_{\mu}^{(a)}
(s)\phi_{s+\hat\mu}^{(a)}
+ h.c)\nonumber\\
&&+\lambda\sum_{a,s}(\phi_s^{(a)\ast}\phi_s^{(a)}-1)^2
+\sum_{a,s}\phi_s^{(a)\ast}\phi_s^{(a)},
\end{eqnarray}
where 
$U^{(a)}=\exp(i\vec{\epsilon}_{a}\cdot\overrightarrow{C})$.\
We modify the partition function of the DGL model
using the Villain approximation $e^{\alpha \cos\psi}\to\sum_{l\in Z}
e^{-\frac{\alpha}{2}(\psi+2\pi l)^2}$.\
In DGL,\ 
the summations of $l^{(a)}$ appear 
due to 2$\pi$ periodicity of 
$\cos(d\varphi^{(a)}+\vec{\epsilon}_{a}
\cdot\overrightarrow{C})$.\
Then there is a constraint $\sum_{a}l^{(a)}=0$,\ since 
$\sum_{a}(d\varphi^{(a)}+\vec{\epsilon}_{a}\cdot
\overrightarrow{C} )=0$.\ The partition function of the DGL 
theory is given by
\begin{eqnarray}
Z_{DGL}&=&\int_{-\infty}^{\infty}{\cal
D}\overrightarrow{C}\int_{-\pi}^{\pi}{\cal
D}\mbox{\boldmath
$\varphi$}\delta(\sum_{a}\varphi^{(a)})\int_0^{\infty}
{\cal D}\mbox{\boldmath $\rho$}^{-2} \sum_{l^{(a)} \in 
Z}\delta_{\Sigma_a l^{(a)},0}\nonumber\\
&&\times\exp\{
-\frac{1}{2g_m^2}||d\overrightarrow{C}||^2
-\gamma\sum_{a,s,\mu}\rho_{s}^{(a)}\rho_{s+\hat\mu}^{(a)}
(d\varphi^{(a)}+\vec{\epsilon}_{a}\cdot\overrightarrow{C}
+2\pi l^{(a)})_{s,\mu}^2 \nonumber\\
&&-\lambda\sum_{a,s}((\rho_{s}^{(a)})^2
-1)^2-\sum_{a,s}(\rho_{s}^{(a)})^2\}.
\end{eqnarray}
Inserting 
\begin{eqnarray}
1&=&\left\{\prod_{a,s,\mu}(4\gamma\rho_{s}^{(a)}
\rho_{s+\hat\mu}^{(a)})^{-1/2}\right\}
\int_{-\infty}^{\infty}{\cal D}\mbox{\boldmath $F$}
\exp\{-\sum_{a,s,\mu}\frac1{4\gamma
\rho_{s}^{(a)}\rho_{s+\hat\mu}^{(a)}}\nonumber\\
&&\times(F_{\mu}^{(a)}(s) -2i\gamma  
\rho_{s}^{(a)}\rho_{s+\hat\mu}^{(a)}(d\varphi^{(a)}
+\vec{\epsilon_a}\cdot\overrightarrow C +2\pi l^{(a)})_{s,\mu})^2 \}
\end{eqnarray}
and using the relation in Eq.\ (5.7),\
we integrate out 
$\phi,\ \varphi^{(a)}$ and $\ \overrightarrow{C}$.\
The partition function is written by the variables $j^{(a)}$ in the
{\it {r.h.s.}}\  of Eq.\ (5.7) and $\rho^{(a)}$.\ 
From the integrals of $\varphi^{(a)}$,\ the relations
$\delta j^{(1)}=\delta j^{(2)}=\delta j^{(3)}$ appear.\ 
Here monopole currents are
introduced as follows: 
\begin{equation}
k^{(1)}\equiv j^{(2)}-j^{(3)},\ k^{(2)}\equiv 
j^{(3)}-j^{(1)},\ k^{(3)}\equiv j^{(1)}-j^{(2)}.
\end{equation}
They satisfy the conservation law $\delta k^{(a)}=0$
and the constraint $\sum_{a}k^{(a)}=0$.\
The partition function can be rewritten as follows:
\begin{equation}
{Z}_{mon}[k^{(a)}]=\sum_{k^{(a)}\in Z}\delta_{\delta
k^{(a)},0}
\delta_{\Sigma_a k^{(a)},0}
\exp \{-S_{mon}^G [k^{(a)}] -S_{mon}^H [k^{(a)}]\},
\end{equation}
\begin{equation}
S_{mon}^G [k^{(a)}]=\frac{g_m^2}{4}\sum_{a=1}^3 
(k^{(a)},\ \Delta^{-1}k^{(a)}),
\end{equation}
\begin{eqnarray}
S_{mon}^H [k^{(a)}]&=&-\ln\{\int_0^{\infty}{\cal D}\mbox{\boldmath $\rho$}^{-2}
\exp\{-\frac{1}{4\gamma}\sum_{s,\mu}\frac{\sum_{a}
\rho_s^{(a)}\rho_{s+\hat\mu}^{(a)}({k}_{\mu}^{(a)}(s))^2}
{\sum_{a < b}\rho_s^{(a)}\rho_{s+\hat\mu}^{(b)}}\nonumber\\
&&-\lambda\sum_{a,s}(({\rho}_s^{(a)})^2 - 1)^2
-\sum_{a,s}({\rho}_s^{(a)})^2\}\}.
\end{eqnarray}
The integration  of  $\rho^{(a)}$ in Eq.\ (A.8) can't be performed exactly.\
In the London limit ($\lambda\to\infty$),\ $\rho^{(a)}$ are fixed to
unity.\ 
$S_{mon}^H [k^{(a)}]$ becomes the quadratic self interaction:
\begin{equation}
S_{mon}^H [k^{(a)}]=\frac{1}{12\gamma}\sum_a ||k^{(a)}||^2.
\end{equation}
The monopole action in Eq.\ (4.1)
\begin{eqnarray}
S_{mon}[k^{(a)}]&=&\sum_{a} (k^{(a)},D_0 k^{(a)}),\ \ \ 
D_0=\alpha \Delta^{-1} +\beta
+\gamma \Delta
\end{eqnarray}
corresponds to the modified London limit of DGL:
\begin{eqnarray}
S_{DGL}[\overrightarrow{C},\varphi^{(a)}]&=&
\frac1{8\alpha}||d\overrightarrow{C}||^2
+\frac1{12}\sum_a (X^{(a)},\{\beta+\gamma\Delta\}^{-1}X^{(a)}),\nonumber\\
&&
X^{(a)}=
d\varphi^{(a)} +{\vec{\epsilon}}_a\cdot\overrightarrow{C}+2\pi l^{(a)}.
\end{eqnarray}
\par
The Wilson loop is also calculated similarly.\ For simplicity,\ 
we consider the case of the London limit.\
Since the DGL theory is the dual theory for a color electric $U(1)^2$ charge,\ 
we consider the 't\ Hooft loop:  
\begin{eqnarray}
\lefteqn{\langle H_c(C) \rangle= \frac1{{Z}_{DGL}}\int_{-\infty}^{\infty}
{\cal D}
\overrightarrow{C}\int_{-\pi}^{\pi}{\cal
D}\mbox{\boldmath $\varphi$}\delta(\sum_{a}
\varphi^{(a)})\sum_{l^{(a)} \in Z}
\delta_{\Sigma_a l^{(a)},0}}\nonumber\\
&&\times\exp\{-\frac{1}{2g_m^2}||d\overrightarrow{C} -
4\pi\overrightarrow{Q}_c{}^* S||^2
-\gamma\sum_a
||d\varphi^{(a)}+\vec{\epsilon}_a\cdot\overrightarrow{C}
+2\pi l^{(a)}||^2\},
\end{eqnarray}
where 
$S$ is a surface with the fixed boundary $C$:\  
$\delta{S}=j$,\ where $j$ is the unit current on the  
loop $C$.\  
$\overrightarrow{Q}_c$ are  color electric $U(1)^2$ charges of 
quarks with  colors c\ (c=R,G,B):
\begin{equation}
\overrightarrow{Q}_c=\left ( Q_c^3,\ Q_c^8 \right ) =
\left\{\left(\frac12,\ \frac1{2\sqrt3}\right),\ \left(-\frac12,\
\frac1{2\sqrt3}
\right),\ \left(0,\ -\frac1{\sqrt3}\right)\right\}.
\end{equation}
For c=R,\ 
we get
\begin{eqnarray}
\lefteqn{\langle H_R(C) \rangle= \frac1{
Z_{mon}}\sum_{k^{(a)} \in Z}\delta_{\delta
k^{(a)},0}\delta_{\Sigma_a k^{(a)},0}
\exp\{-\frac{g_m^2}{4}\sum_{a}
(k^{(a)},\ \Delta^{-1}k^{(a)} )}\nonumber\\
&&-\frac1{12\gamma}\sum_{a}||k^{(a)}||^2
+2\pi i({}^* S,\ \Delta^{-1}dk^{(3)})
-\frac{8\pi^2}{3g_m^2}({}^*j,\ \Delta^{-1}{}^*j)\}.
\end{eqnarray}
The monopole Wilson loop operators are
\begin{equation}
e^{2\pi i({}^* S,\ \Delta^{-1}dk^{(3)})},\ \ 
e^{2\pi i({}^* S,\ \Delta^{-1}dk^{(2)})},\ \ 
e^{2\pi i({}^* S,\ \Delta^{-1}dk^{(1)})}
\end{equation}
for c=R,G,B,\ respectively.\
When ${}^* S$ is a closed surface,\ 
$({}^* S,\ \Delta^{-1}dk^{(a)})$ in Eq.\ (A.15) is the
4-dimensional linking number
between the closed surface ${}^* S$ and the closed loop
$k^{(a)}$.\ 
Hence the monopole Wilson loop is independent of the
choice of the surface $S$.\
This is very important.\par
Let us consider the string representation of
Eq.\ (A.12).\
Introducing the following $U(1)^2$ electric charges:
\begin{equation}
\mbox{\boldmath $s$}_c = (s_c^{(1)},s_c^{(2)},s_c^{(3)})=\left\{
(1,-1,0),(-1,0,1),(0,1,-1)\right\}
\end{equation}
for c=R,G,B respectively [36],\ the string representation
can be derived as in Section 5:
\begin{eqnarray}
&&\langle H_c(C)\rangle=\frac1{
Z_{str}}\sum_{\sigma_c^{(a)}
\in Z,\delta{}^*\sigma_c^{(a)}=j_c^{(a)}}
\delta_{\Sigma_a \sigma_c^{(a)},0}
\exp\{-4\pi^2\gamma \{\sum_{a}(\sigma_c^{(a)},
\ (\Delta+m^2)^{-1}\sigma_c^{(a)})\nonumber\\
&&\ \ \ \ \ \ \ \ \ \ \ \ \ \ \ \ \ \ \ \ \ \ \ \ \  \ \ \ \ \ \ \ \ \ 
\ \ \
+\frac1{m^2}\sum_{a}(j_c^{(a)},\
(\Delta+m^2)^{-1}j_c^{(a)})\}\},
\end{eqnarray}
where $j_c^{(a)} \equiv s_c^{(a)} j$ are electric $U(1)^2$
currents and $m^2=3\gamma g_m^2$ is a mass of a dual gauge field in DGL.\
The two-forms $\sigma_c^{(a)}\equiv s_c^{(a)} S+\sigma^{(a)}$
satisfy $\delta{}^*\sigma_c^{(a)}=j_c^{(a)}$ with the constraint
$\sum_a\sigma_c^{(a)} =0$.\ 
Thus we obtain the hadronic string model in the SU(3) lattice QCD.\

\end{document}